\documentclass[a4paper,twoside]{article}
\pdfoutput=1
\newcommand{\affil}[1]{$^{\rm #1}$}
\usepackage[authoryear]{natbib}
\bibpunct{(}{)}{;}{a}{}{,}
\usepackage{graphicx}
\usepackage{xspace}
\usepackage{amsmath}
\usepackage{amssymb}
\date{} %Please leave the date blank
\newcommand{\duchamp}{Duchamp\xspace}
\newcommand{\selavy}{Selavy\xspace}
\newcommand{\eg}{e.g.\xspace}
\newcommand{\ie}{i.e.\xspace}
\title{\large\bf\flushleft Source-finding for the Australian Square Kilometre Array Pathfinder}
\author{\parbox{\textwidth}{\flushleft
\vspace{-0.5cm}
{\it Matthew Whiting\affil{A,B} and Ben Humphreys \affil{A}}\\
\vspace{0.4cm}
{\small \affil{A}\,CSIRO Astronomy and Space Science, P.O. Box 76,
  Epping, NSW, 1710, AUSTRALIA}\\
{\small \affil{B}\,Email: Matthew.Whiting@csiro.au}}}
\begin{document}
\twocolumn[
\begin{changemargin}{.8cm}{.5cm}
\begin{minipage}{.9\textwidth}
\vspace{-1cm}
\maketitle
%
%
%%%%%%%%%%%%%     ABSTRACT    %%%%%%%%%%%%%
%Abstract of no more than 200 words here.
\small{\bf Abstract:}
The Australian Square Kilometre Array Pathfinder (ASKAP) presents a
number of challenges in the area of source finding and
cataloguing. The data rates and image sizes are very large, and
require automated processing in a high-performance computing
environment. This requires development of new tools, that are able to
operate in such an environment and can reliably handle large
datasets. These tools must also be able to accommodate the different
types of observations ASKAP will make: continuum imaging,
spectral-line imaging, transient imaging. The ASKAP project has
developed a source-finder known as \selavy, built upon the \duchamp
source-finder \citep{whiting12}. \selavy incorporates a number of new
features, which we describe here.

Since distributed processing of large images and cubes will be
essential, we describe the algorithms used to distribute the data,
find an appropriate threshold and search to that threshold and form
the final source catalogue. We describe the algorithm used to define a
varying threshold that responds to the local, rather than global,
noise conditions, and provide examples of its use. And we discuss the
approach used to apply two-dimensional fits to detected sources,
enabling more accurate parameterisation. These new features are
compared for timing performance, where we show that their impact on
the pipeline processing will be small, providing room for enhanced
algorithms.

We also discuss the development process for ASKAP source finding
software. By the time of ASKAP operations, the ASKAP science
community, through the Survey Science Projects, will have contributed
important elements of the source finding pipeline, and the mechanisms
in which this will be done are presented.

%%%%%%%%%%%%%     KEYWORDS    %%%%%%%%%%%%%
\medskip{\bf Keywords:} methods: data analysis --- methods: numerical
--- techniques: image processing
% Please write all keywords in lower case. PASA uses the
% standard list of subject headings adopted by The Astrophysical Journal
% and available from http://www.journals.uchicago.edu/ApJ/keywords_text.html.
% Keywords are separated by em-dashes, i.e. ---

%%%%%%%%DO NOT EDIT%%%%%%%%%%%%
\medskip
\medskip
\end{minipage}
\end{changemargin}
]
\small
%%%%%%%%EDIT FROM HERE%%%%%%%%%%%%

\section{A source-finder for ASKAP}

The Australian Square Kilometre Array Pathfinder \citep{deboer09} is
an aperture synthesis radio telescope currently under construction in
the radio-quiet Western Australian outback. It will be an array of 36
antennas, each equipped with a focal plane phased-array feed (PAF),
operating between 700\,MHz and 1.8 GHz and capable of a field-of-view
of 30 square degrees.  The first five years of ASKAP operations will
have at least 75\% of the time dedicated to large Survey Science
Projects (SSPs), each requiring more than 1,500 hours. Ten SSPs have
been identified, many of which require the formation of large (often
all-sky) source catalogues.

ASKAP observations will produce very large data rates, as a result of
the large number of beams that give the wide field-of-view, the large
number of baselines, the large instantaneous bandwidth and spectral
resolution (300\, MHz divided into 16384 channels), together with
5-second sampling.  At full spectral resolution, the visibility
dataset will be $\sim80$ terabytes after 8 hours of observing, and
will be reduced to image cubes of $\sim 1$ terabyte that need to be
searched for astronomical sources. These data rates necessitate
processing in automated pipelines running on a highly distributed
parallel processing computer. They also force the adoption of
particular algorithmic choices in the imaging, and resource
availability will inevitably lead to limitations in the processing
capabilities (For instance, the quoted value of $\sim 1$ terabyte for
a cube is for a spatial resolution of 30$^{\prime\prime}$. This is the
highest possible for the full spectral resolution imaging, but
low-resolution continuum imaging will be possible at
$\sim10^{\prime\prime}$). 

The software pipelines for ASKAP are currently under development, but
prototypes exist of all the main elements, from ingest of visibility
data, calibration, imaging and source extraction. These pipelines will
be used to process data from BETA, the Boolardy Engineering Test
Array, which consists of the first six ASKAP antennas equipped with
phased-array feeds. BETA will have the same field of view as ASKAP,
but with much coarser spatial resolution, meaning the number of
spatial pixels required will be smaller. However, BETA will have the
same number of spectral channels as ASKAP, and so the spectral-line
images produced by the pipelines will be comparable in size (despite
the visibility data sets being smaller by virtue of the reduced number
of baselines).

This paper focusses on the last element of the pipeline processing --
the source extraction. This pipeline is built on the software library
of the stand-alone \duchamp source finder \citep{whiting12}, and adds
features not included in \duchamp. This paper describes the
development process, and details some of the new features that have
been implemented.

We give a brief description of \duchamp here, but readers are directed
to \citet{whiting12} or the Duchamp User's Guide for specific details
about the \duchamp algorithms themselves.

\duchamp is a standalone program, developed at CSIRO Astronomy \&
Space Science independently of the ASKAP project, and publicly
available\footnote{http://www.atnf.csiro.au/people/Matthew.Whiting/Duchamp}. It
was developed primarily to find sources in three-dimensional
spectral-line data cubes, although it is able to process two- and
one-dimensional data as well. One of its key features is the ability
to pre-process the image data via smoothing or multi-resolution
wavelet reconstruction to minimise the effects of noise and increase
both the completeness and reliability of the resulting source
catalogue.

\duchamp, however, lacks certain features that would make it suitable
for ASKAP online processing, in particular parallel processing of
data. It also lacks certain features that ASKAP surveys would
desire. We have been developing an ASKAP source-finder that builds on
the \duchamp library, extending it in appropriate areas. The current
version of this source-finder is known as \selavy\footnote{Rrose Selavy
  was a pseudonym of Marcel Duchamp, after whom \duchamp was named.}

\selavy was developed specifically to operate in a distributed
environment, and also features improvements to the detection and
parameterisation algorithms. These innovations are detailed in the
following sections: Section~\ref{sec-distrib} describes how the
software is adapted to work in a distributed environment;
Section~\ref{sec-thresh} describes changes to the determination of the
threshold, allowing it to operate in a distributed environment and
allowing the threshold to vary from pixel to pixel; and
Section~\ref{sec-fit} describes additional processing enabled for
continuum images, that allow better analysis of detected
sources. Finally, Section~\ref{sec-devel} describes the development
process, and how interactions with the community are aiding the
development.

\section{Distributed Processing}
\label{sec-distrib}

\subsection{Why distributed processing?}

The large field of view and spectral coverage of ASKAP place great
demands on the processing capability, driving us towards distributed
processing. The size of the ASKAP spectral-line data sets demands it,
as a full cube (typical size $3600\times3600$ pix $\times 16384$
channels, or nearly 800GB) will not fit in memory for a single
processor. The ASKAP continuum data, being single-channel images (at
least, the images that result from the multi-frequency synthesis
imaging), will easily fit in memory, but the large field of view
results in such a large number of sources (\citet{norris11} predict
the EMU survey will find $\sim70,000$ sources per ASKAP field) that
parallel processing is required to meet the performance goals of the
pipeline. 

In general, splitting up the data set allows it to be processed in
parallel, decreasing the processing time and potentially allowing a
number of different approaches, or more computationally-intensive
analyses to be used. Finally, the ASKAP pipeline processing will take
place within a high-performance supercomputing environment (this is
driven more by the imaging, which has even stricter requirements on
distributed processing), and so distributed source extraction will
make the best use of the available resources.

We describe in this section the framework and implementation that is
being developed for the ASKAP pipeline source finder. This has been
tested on large multi-core machines such as the
NCI\footnote{http://www.nci.org.au} National Facility Sun
Constellation \textit{vayu} and the iVEC\footnote{http://www.ivec.org}
Pawsey 1A machine \textit{epic} -- the latter being the machine that
will be used for processing data from BETA -- and is the subject of
on-going development and evaluation.

\subsection{Implementation}

The systems the source-finding runs on are characterised by having a
large number of nodes, each comprising typically 8-12 CPU cores with
an average of 2-3GB of memory per core. We therefore cannot assume
that the entire image will fit within memory. The source-finding
implements the master/worker pattern for workload distribution and
coordination. A single processing task is run using $N$ processes, one
of which is designated the "master" process and coordinates the
processing, and the remaining $N - 1$ are "worker" processes and perform
the compute intensive work. The master/worker pattern has proved to be
more then adequate to meet the scaling goals, partially owing to the
fact that work tasks are loosely coupled and relatively
coarse-grained, requiring significant CPU time to complete.

The first step in the distributed processing is to allocate a subimage
of the full input image to a given worker. This is done by dividing
the image at regularly-spaced intervals in each axis direction, via
parameters $N_x$, $N_y$ and $N_z$ (where the x- and y-directions are
the spatial directions, such as right ascension and declination, and
the z-direction is the spectral direction, such as frequency or
velocity). The number of workers required is thus given by the product
of these three parameters: $N-1 = N_x N_y N_z$. The size of these
worker images can be made larger than necessary by allowing them to
overlap by some specified amount. This allows sources at or near the
edge to be better recovered (although see
Section~\ref{sec-edgesources} for further discussion on these
sources), and allows complete coverage of the image when using the
variable threshold technique discussed in Section~\ref{sec-varthresh}.

Each worker reads its subimage data from disk independently, then
processes its own subimage using \duchamp algorithms as well as those
described herein, constructing a list of objects to be sent to the
master. The worker only sees the pixels within the subimage, although
it has information about where that subimage is located within the
complete input image, so that correct pixel locations can be assigned
(rather than just the pixel locations within the subimage).

\subsection{Sources at subimage edges}
\label{sec-edgesources}

When dividing up an image for processing, one needs to consider the
effect of the edges of the subimages. It is likely that these will lie
on or near sources of interest, and so care must be taken to ensure
these sources are not processed differently to sources away from the
edges (since the edges are arbitrary and not related to the data
itself). 

For sources away from the edges of the subimages, the processing is
identical to the single-threaded case. The worker has all the flux
information for every pixel in the source, and so the source can be
fully parameterised. Any fitting (described in Section~\ref{sec-fit}) is
done by the worker as well. The only additional impact is the work
involved in sending the information on the source to the master node,
which then writes it to the output.

When a source is close to the edge of a subimage, however, it is
likely that the entire source does not lie in a single subimage, and
so a single worker cannot completely process the source. These sources
are flagged as edge sources and are handled by the master differently
to the non-edge sources. All edge sources are compiled into a list,
and then passed through the same merging process used by \duchamp (see
\citet{whiting12}). This provides a list of unique sources, distinct
from the non-edge sources.

Before the lists of edge and non-edge sources can be combined,
however, the edge sources need to be parameterised.  The sources are
distributed to the now-idle workers, who do the basic parameterisation
of each source individually. This is currently handled by the \duchamp
algorithms, which limit the parameterisation (object extent, peak and
integrated flux) to the detected pixels only. A worker will only need
the pixels that immediately affect a given source, and so will not
need to load the full image, or even a full subimage. This will keep
the memory requirements tractable (we assume the sizes of the objects
are small compared to the size of the cube).

\section{Threshold determination in \selavy}
\label{sec-thresh}

The \duchamp software uses a single threshold, either given as a flux
value or a signal-to-noise level, for the entire dataset that is being
searched. This approach results in an output catalogue with a uniform
selection criterion. It is best suited to data that has uniform noise,
which often requires some form of preprocessing (for instance,
division through by the sensitivity pattern). In practice, for ASKAP
processing, there are at least two issues with this approach. One is
that the image data is distributed, so that no single worker process
has access to all the pixel values. Secondly, it is likely that the
assumption of uniform noise everywhere will not hold. While it may be
possible to divide through by the ASKAP sensitivity pattern,
additional effects such as sidelobes will contribute to the noise in
different ways at different locations In this section, we discuss two
approaches \selavy can implement to address these issues.

\subsection{Statistics in the distributed case}
\label{sec-stats}

\begin{table*}[ht]
    \begin{center}
      \caption{Image statistics in distributed processing.}
      \label{tab-stats}
      \begin{tabular}{lcc|cc}
        \hline \# Workers & $m_M$& $s_M$  &$m_M$ &$s_M$\\
        &\multicolumn{2}{c|}{noise + sources} &\multicolumn{2}{c}{noise only} \\
        \hline
        1 & -5.0509e-06 & 3.3746e-05& -1.9926e-07 & 7.8009e-04\\
        2 & -5.0929e-06 & 3.3807e-05&-2.0099e-07 & 7.8025e-04\\
        4 & -5.1624e-06 & 3.4080e-05&-1.5316e-07 & 7.8024e-04\\
        9 & -5.2095e-06 & 3.4454e-05&-2.1657e-07 & 7.8028e-04\\
        16 & -5.2548e-06 &3.4394e-05&-1.0837e-07 & 7.8282e-04\\
        30 & -5.3137e-06 & 3.4641e-05&-1.7546e-07 & 7.8485e-04\\
        \hline
        1$^a$&  1.5374e-05 & 3.0777e-03 &2.3859e-07 & 8.0582e-04\\
        \hline
      \end{tabular}
      \medskip\\
      $^a$ Calculated with non-robust statistics.\\
      $m_M$ = estimate of overall mean, calculated by the master process\\
      $s_M$ = estimate of overall standard deviation, calculated by the master process\\
    \end{center}
\end{table*}

A signal-to-noise threshold is defined by measuring the image
statistics and setting the flux threshold to be a certain number of
standard deviations above the mean. The mean $m$ and standard
deviation $s$ of the pixel values are either measured directly via the
standard relations
\[
m=\frac{1}{N}\sum_{i=1}^N F_i
\]
\[
s=\sqrt{\frac{1}{N}\sum_{i=1}^N(F_i-m)^2}
\]
(where $F_i$ are the pixel flux values), or estimated robustly using
the median and the median absolute deviation from the median (MADFM)
\[
m=\operatorname{med}(F)
\]
\[
s = \operatorname{med}\left(|F - m|\right) \times 0.6744888
\]
(where the correction factor 0.6744888 converts the MADFM to the
equivalent standard deviation of a Normal distribution
\citep{whiting12}). The robust methods avoid the strong bias that can
be present from the inclusion of source pixels (which do not form part
of the noise background anyway), albeit at the expense of additional
computation.

If one wishes to apply a single signal-to-noise threshold for the
entire image, in the fashion of \duchamp, then the global image
statistics need to be known. In the distributed processing case, no
single worker process can see the entire image, and having each worker
calculate their own statistics would lead to large-scale
discontinuities in the detection threshold. There must, instead, be a
way of estimating the global statistics from the noise properties of
the individual images.

First, each worker finds the mean or median of their subimage, and
sends this to the master. The master then forms a weighted average of
the workers' means (weighting by the number of pixels in each
subimage):
\[
m_M = \frac{\sum_i m_i N_i}{\sum_i N_i}
\]
(here the subscript $M$ refers to the value at the master, while $i$
refers to a given worker). This provides an estimate of the global
mean (in fact, when means are used instead of medians, this is
identical to the overall mean).

This mean is then distributed to the workers, who use it to find the
``spread'' (either the standard deviation or the MADFM) within their
subimage and provide this to the master. The master then forms a
weighted average of the workers' variances
\[
s_M = \sqrt{\frac{\sum_i s^2_i N_i}{\sum_i N_i}}
\]

When using the mean and standard deviation, these quantities come out
to the same values one would get by looking at the entire image at
once. However, the robust statistics will be different. It is not
possible to find the overall median without looking at all the data
(since at least partial sorting of the entire dataset would be
required), but taking the weighted average of the medians provides a
better estimate than the median of the medians.

The accuracy of the distributed approach is affected by the number of
workers used. In Table~\ref{tab-stats} we look at how the estimates of
the overall image statistics, estimated with the robust methods
described above, change with the number of workers used. This is done
for two cases, taken from the 2011 ASKAP simulations: one, a continuum
image that has many sources present, with signal-to-noise values
ranging up to $\sim10^5$; and two, a single channel of a
noise-dominated spectral cube. In both cases, the estimate of the
standard deviation increases with the number of workers, and increases
more strongly in the source-dominated case. Even though we are using
robust statistics, the estimates in subimages with bright sources are
still slightly affected, and their inclusion increases the calculated
global value. Even though there are no sources in the second case, the
estimated noise still increases, albeit by only a tiny amount. Note
that using non-robust statistics in the source-dominated image gives a
standard deviation almost two orders of magnitude larger than the
robust statistics.

These considerations are important only in the case of determining the
global noise properties, or in applying a global signal-to-noise
threshold. In Section~\ref{sec-varthresh} we discuss an alternative
method of setting the threshold that depends only on the \emph{local}
noise properties.

\subsection{Variable thresholds}
\label{sec-varthresh}

\begin{figure*}
  \begin{minipage}{\textwidth}
    \includegraphics[width=\textwidth]{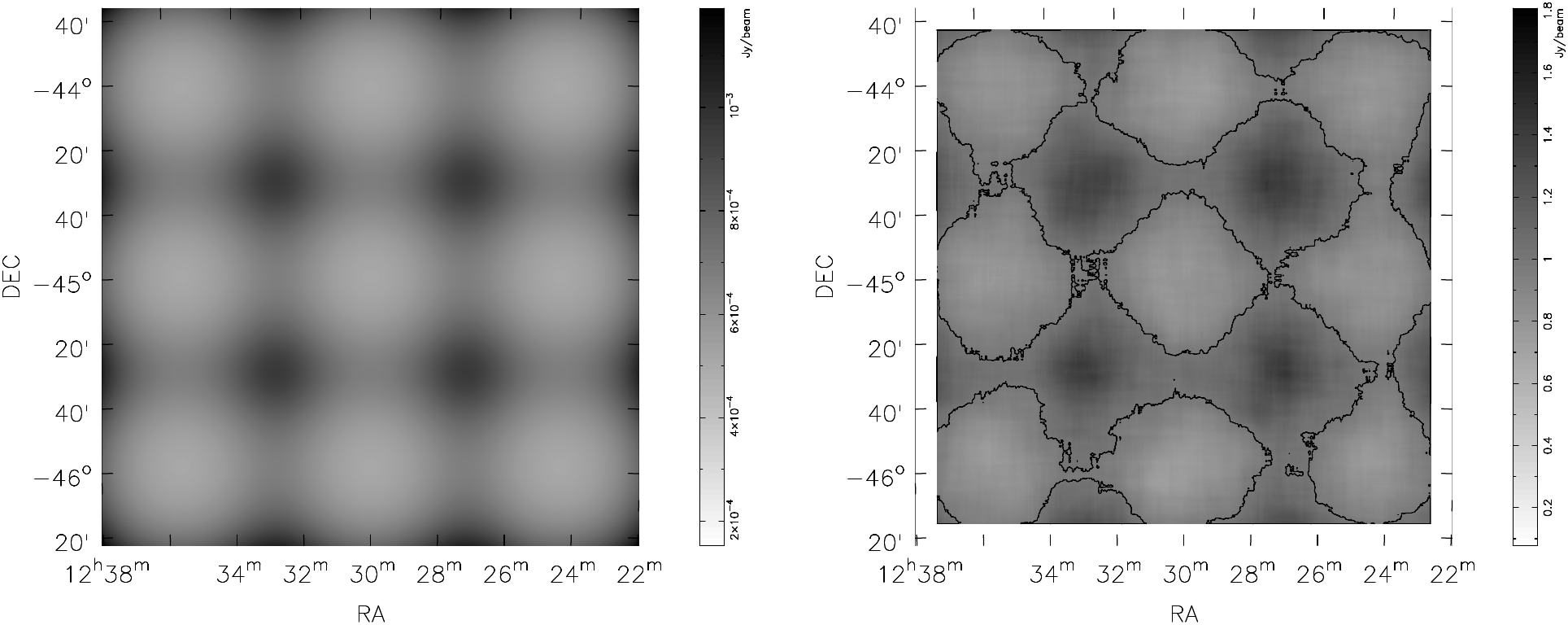}
    \caption{An illustration of the effect of applying the variable
      threshold technique in the presence of sensitivity
      variations. The left-hand plot shows the variations in
      sensitivity (\ie the theoretical noise level) due to a 3x3 grid
      of beams, taken from a simulated ASKAP observation. White is
      more sensitive. The right-hand side shows the result of the
      variable threshold determination on a particular noisy image
      with this sensitivity pattern. The image shows the ratio of the
      variable threshold (calculated as described in the text, with a
      box width of 101 pixels) to the single threshold determined from
      the whole image. A $4\sigma$ threshold was used. The single
      contour line indicates where the two thresholds are
      equal. Darker pixels have a higher variable threshold, and
      lighter pixels have lower. }
    \label{fig-varThreshSens}
  \end{minipage}
\end{figure*}

\begin{figure*}
  \begin{minipage}{\textwidth}
    \includegraphics[width=\textwidth]{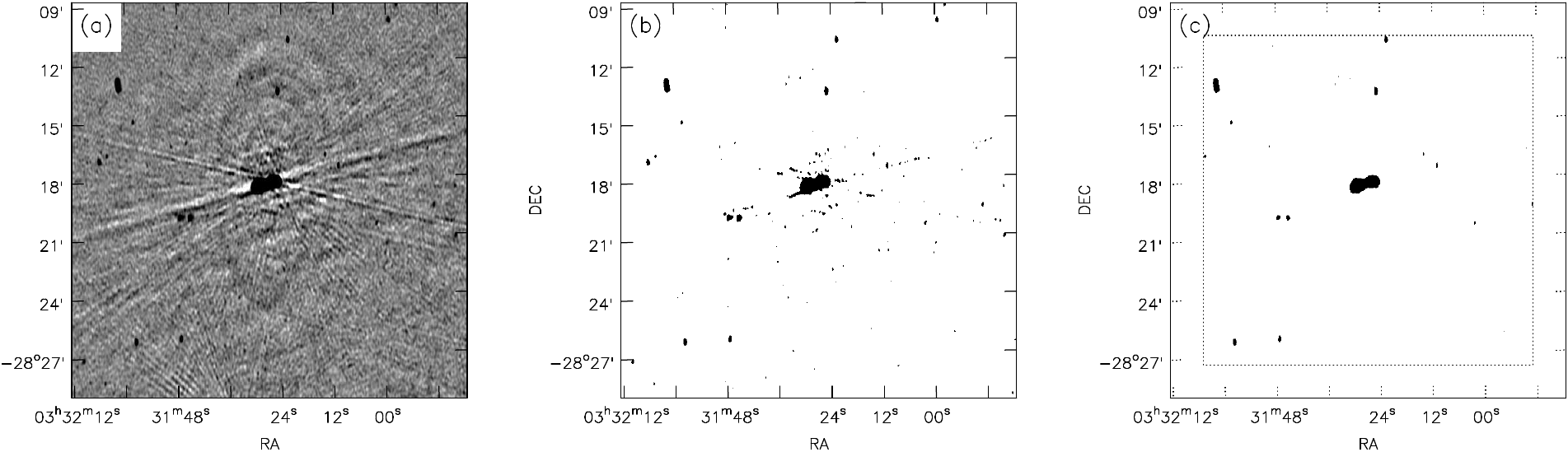}
    \caption{An illustration of the effect of applying the variable
      threshold technique in a situation where sidelobes are a
      problem. Panel (a) shows an excerpt from the ATLAS CDFS field
      \citep{norris06}, showing a bright source surrounded by a
      sidelobe pattern as well as fainter real sources. Panel (b)
      shows the mask resulting from source detection done with a
      constant $5\sigma$ threshold, while panel (c) shows the effect
      of a $5\sigma$ threshold applied using the local noise
      technique. The dotted line indicates the border of the valid
      area - pixels outside this will not be detected.}
    \label{fig-varThreshSidelobe}
  \end{minipage}
\end{figure*}

\subsubsection{Rationale}

One of the key aims of source extraction is to provide a catalogue
that is as complete and reliable as possible. To do this, one needs to
go as deep as possible (to increase the completeness), but not so deep
that one finds large numbers of false detections (which would reduce
the reliability). If the noise varies as a function of position, then
the detection threshold should also vary. By 'noise', here, we mean
not just the random, thermal noise that is inevitably a part of any
image, but also non-random, unwanted signal such as sidelobes or
interference. Thus, at locations where there is a relatively large
amount of noise or additional signal (\eg near the edge of the field
or near a bright source), the threshold can be set higher to avoid
spurious sources, but at other locations it can be set as low as the
thermal noise permits, allowing the source extraction to be as
complete as possible.

An important application of a variable detection threshold is for
non-interlaced observations with ASKAP. The phased-array beams will be
separated by $(\lambda/D)$, rather than the $(\lambda/2D)$ required to
image without aliasing. This leads to variations of $\sim20-30$\% in
the sensitivity across the field. To maximise the completeness and
reliability of any catalogue, the detection threshold must be able to
track these variations. Note that interlacing multiple observations
will provide a flat sensitivity response, and this will be considered
in planning ASKAP surveys.

\subsubsection{Implementation}
\label{sec-varThreshImpl}

\selavy approaches the goal of maximising the completeness and
reliability by finding, for each pixel, the image statistics within a
'box' of a given size centred on that pixel. The image statistics are
used to set a signal-to-noise threshold, which then determines whether
that pixel is part of an object or not. This is repeated for each
pixel in the image, thereby providing a different flux threshold at
each location.

The implementation of this uses the \textsc{casacore} library's
\textit{slidingArrayMath} functions, which allow efficient sliding of
a box over an array. Although described as ``boxes'', these can be
either two- or one-dimensional, to match the method of searching done
by the \duchamp algorithms (which search a 3D dataset either one
channel image at a time or one spectrum at a time).

Applying a fixed box size to the array means that pixels within half
the box width of the edge cannot have a full box centred on them
without it extending past the image borders. These pixels have their
signal-to-noise ratio set to zero, and so will yield no
detections. This has implications for the distributed processing
discussed in Section~\ref{sec-distrib}, but problems can be
avoided by using an overlap between neighbouring worker subimages of
at least the box width. The edge of the image, however, will always
have a border area that exhibits no detections.

The box size should be chosen carefully. If it is too small, a source
may fill a large fraction of the box and so the noise estimate will
not sample the true background. If it is too large, any sensitivity
variations present (see Fig.~\ref{fig-varThreshSens}) may get smoothed
out and the utility of the approach diminished.

At present, only a single box size is applied to the data, but there
is the risk that this may impose a preferred scale on the output
catalogue, particularly if there is underlying large-scale diffuse
structure in the image that may be comparable in size to the box. An
improved algorithm would make use of a range of box sizes and
appropriately account for the different detection thresholds that
would result -- this is an area of ongoing research within the ASKAP
source-finding community.

\subsubsection{Examples}

We consider here two different situations where applying this
technique can be beneficial. Fig.~\ref{fig-varThreshSens} shows how
sensitivity variations can be accounted for in an ASKAP image. The
left-hand image shows a sensitivity pattern taken from an ASKAP
simulation\footnote{These simulations were provided by the ASKAP team,
  and are available from
  http://www.atnf.csiro.au/people/Matthew.Whiting.},
in this case, a single channel from the spectral-line simulation. The
lighter areas are the PAF beams, where the noise (indicated by the
colour scale) is lowest, with the darker areas the increased
noise in between beams. These areas of increased noise are more likely
to contribute spurious sources, particularly at lower (\ie more
interesting) detection thresholds. 

The right-hand image shows the ratio of the variable flux threshold at
each pixel, determined as in Sec.~\ref{sec-varThreshImpl} using a
$4\sigma$ signal-to-noise threshold, to the single $4\sigma$ threshold
(as a flux value) determined from the entire image. A single contour
line marks the locations where the thresholds are equal. One clearly
sees that the flux threshold now tracks the noise variations closely
(note that the left-hand side image shows the \emph{theoretical}
noise, whereas the right-hand side reflects a particular instance of
the random noise ), which will increase the number of real sources
detected (in the PAF beam directions) and decrease the number of false
detections in the higher-noise regions between the beams.

Fig.~\ref{fig-varThreshSidelobe}, meanwhile, shows the effect of
applying this variable threshold approach to data from the Australia
Telescope Large Area Survey \citep[ATLAS,][]{norris06}. We use the
image of the Chandra Deep Field South (CDFS), as it provides a good
illustrative example of sources with strong sidelobes. (Although note
that the sidelobes in this Australia Telescope Compact Array image are
much more prominent than any expected in ASKAP images, due both to the
design of the ASKAP array, with many more baselines and 3rd-axis
rotation of the feed, and to the use of a sky model in the
imaging. This example can be considered a worse-case scenario for
ASKAP imaging.) When we apply a single threshold to this image (with
the noise determined from a part of the image away from bright
sources), these sidelobes appear either as separate detections or
extensions to the primary object (see
Fig.~\ref{fig-varThreshSidelobe}b). Raising the threshold around this
object means that we detect just the central part, but we still detect
the faint sources in the field where the detection threshold remains
low (Fig.~\ref{fig-varThreshSidelobe}c). \citet{huynh12} have made a
detailed examination of this algorithm in looking at source extraction
from ASKAP simulations.

\section{Two-dimensional source fitting}
\label{sec-fit}

\subsection{Motivation}

The principle aim of the \duchamp source-finder is to locate sources
of interest within an image. It makes no assumptions as to the nature
of the sources themselves, and so does not perform any fitting to the
detected sources to do parameterisation. Instead, all parameterisation
is done solely from the pixels in the image.

The rationale here is that the source finding segments the image into
'object' pixels and 'background' pixels, and that the objects of
interest are, by definition, made up of the detected pixels, so
parameterising them by those pixels should be sufficient. In the
absence of any assumption about their true nature, this is all that
can be done. \duchamp then leaves the analysis here, and provides the
user with enough information (such as mask images) to go and do
further, more detailed parameterisation (\eg through fitting)
according to the science they want to do.

In practice, however, there are often assumptions that can be made
about the nature of the sources being detected. A common one in radio
imaging is that the sources' spatial structure can be decomposed into
a small number of Gaussian components, particularly when they are
unresolved (or only marginally resolved). This has been done with many
continuum surveys such as NVSS \citep{condon98}, FIRST
\citep{becker95,white97} and SUMSS \citep{mauch03}, and in many
spectral-line surveys, such as HIPASS \citep{barnes01}, where the
fitting is done on the moment-0 map of a spectral-line source.

The key to this approach is to represent the sky accurately with a
minimal number of easily-quantifiable components. This will facilitate
the cataloguing of the image (as each component can be readily
expressed as a single catalogue entry), and makes the parametrisation
of sources an easier task as well. The Gaussian shape, moreover,
closely approximates the response of radio interferometers, either
after deconvolution, or, in the case of good $u-v$ coverage (such as
long integrations with ASKAP), even before deconvolution, and so
provides a natural basis for representing the image brightness.

We have therefore implemented two-dimensional Gaussian fitting in the
ASKAP source finder, to act as the parameterisation step following the
identification of sources. In the following section we describe the
implementation, with particular emphasis on how to run this within the
ASKAP pipeline environment. The details of the implementation are, at
this point, not final, and are most likely not yet the optimal
solution. Testing is on-going -- see, for instance \citet{hancock12},
who present a very promising alternative algorithm -- and the final
version of the ASKAP pipeline will depend strongly on community input
(see Section~\ref{sec-devel}).

\subsection{Fitting algorithm}

The Gaussian fitting routine starts with the result of the source
finding. This provides, for a given object, a set of pixels (commonly
referred to as an 'island') that meet the detection criteria. These
will be surrounded by background pixels that are, by definition, not
part of the source. 

The fitting can be done in one of two ways. Either just the detected
pixels (their locations and fluxes) are passed to the fitting
algorithm, or all the pixels within a rectangular box surrounding the
object (padded out by some pre-defined number of pixels) are used
instead. 

The former is preferred, as the fitting then is only constrained by
the pixels known to make up the object. This does, however, require a
certain number of pixels to be detected for the fit to be reliable -- a
source that has only the top few pixels of its profile detected may
not have enough pixels for the fit to be constrained, and even if it
does it may not provide a good estimate of the position and shape of
the Gaussian. For this reason, extending detections out to some
secondary flux threshold (known in \duchamp as ``growing'') is used to
provide as much information on the source as possible.

The alternative method of fitting within a box includes all pixels
without applying this secondary threshold, and so hopefully includes
in the fitting all pixels (with significance below the detection
threshold) that contribute to the source. 
The downside is that if neighbouring
sources encroach into the box, without merging with the source under
consideration, then they will also affect the result of the fit, and
may end up having components erroneously fitted to them.

\subsection{Initial guesses}

For accurate results, most fitting algorithms (including the
\textsc{casacore} algorithms used by \selavy) benefit from a good
\textit{a priori} guess for the result. This allows the optimisation
to converge to the global minimum $\chi^2$ value, rather than some
local, but not global, minimum. This is particularly important when
fitting to confused or merged sources - that is, the island of pixels
comprises several components that are joined at some flux level. If
the algorithm can start with relatively accurate guesses of the
location and size of the Gaussian components present then it will
converge faster and more accurately. 

\begin{figure}[t]
  \includegraphics[width=0.5\textwidth]{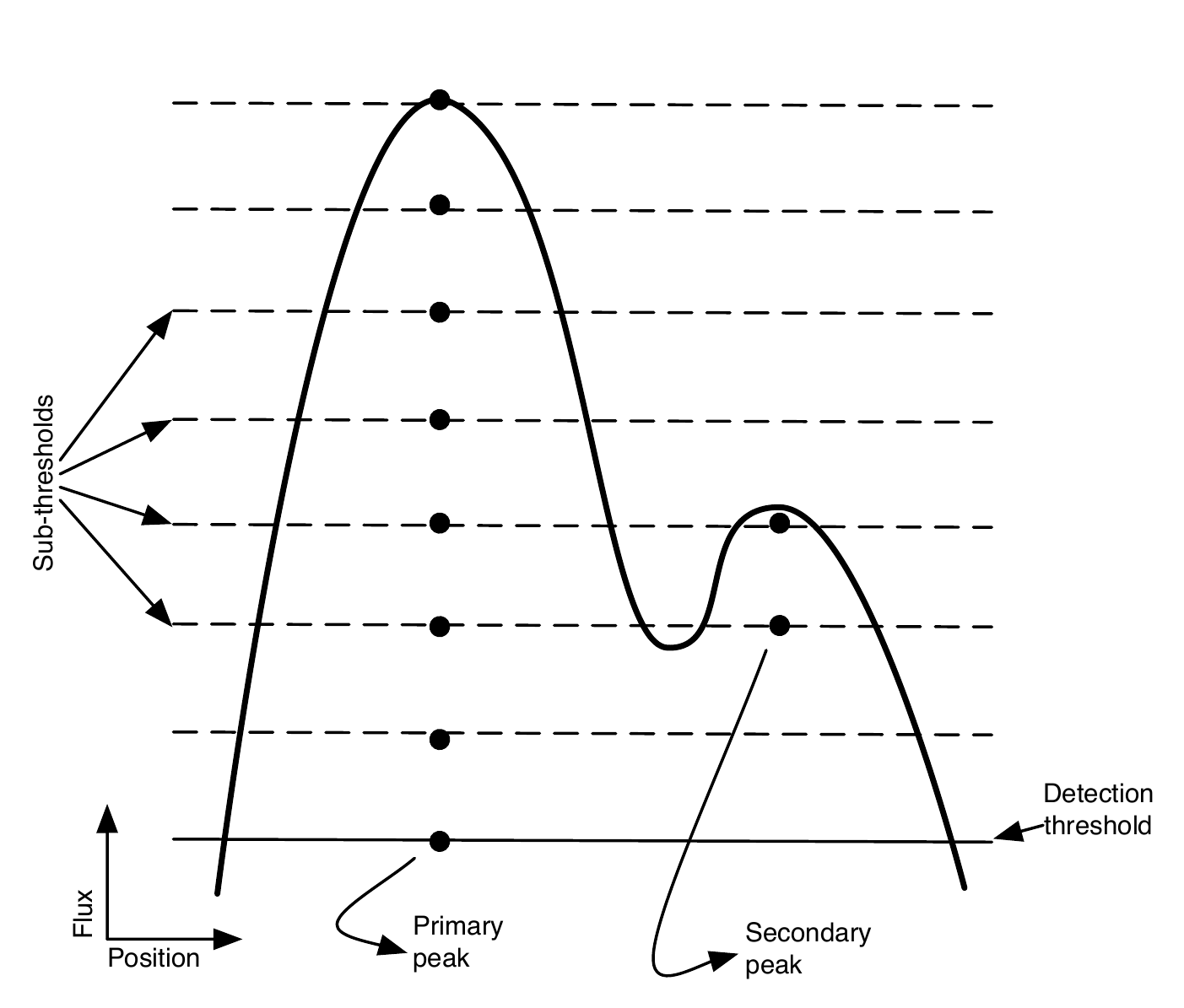}
  \caption{An illustration of the sub-thresholding approach used to
    obtain initial guesses for the Gaussian components present. The
    figure shows a simplified one-dimensional source for clarity. A
    series of subthresholds (dashed lines) are applied between the
    source's peak and its detection threshold. The location of each
    distinct peak is indicated by the filled circle, and would be
    recorded as a separate subcomponent.}
  \label{fig-deblendingexample}
\end{figure}

The initial guesses are determined by a process of sub-thresholding,
illustrated by Fig.~\ref{fig-deblendingexample}. This figure shows a
one-dimension representation of an object comprising two
partially-merged components, both of which peak above the
threshold. The algorithm starts with the island of detected pixels,
and applies a series of thresholds spaced evenly between the
source-detection threshold and the peak pixel (the spacing can be even
in either linear- or log-space). At each threshold, simple source
extraction is done to find the number of components. Each component is
referenced by its peak location, which will remain constant for
different thresholds. If the source has just one component, each
sub-threshold will also return a single source. If there is a
secondary component that is sufficiently well separated, then there
should be a sub-threshold that separates them and returns two
components. The location of each of these are recorded (based on the
peak pixel, so that different sampling of a source does not affect its
location) and will provide a separate initial component to the fitting
algorithm.

The drawbacks of this approach are if the second component is not
sufficiently separated from the primary to provide a 'dip' in flux
between the two. If this is the case, no sub-threshold will be able to
separate them. It is also highly dependent on the specific
sub-thresholds used, such that the sub-threshold increment is able to
resolve the gap between the peak and trough of the secondary
component. 

The Aegean source-finder \citep{hancock12} has an alternative method
of finding subcomponents. It uses a Laplacian filter to construct a
curvature map, searching for local maxima. This algorithm is under
consideration for inclusion in the ASKAP pipeline.

\subsection{Accepting the fit}
 
A given fit is primarily judged as acceptable based on its
goodness-of-fit measure, the $\chi^2$ value. This is the parameter
minimised in the fitting procedure. However, other factors are taken
into account in accepting a fit (these largely follow the procedure of
\citet{white97} for the FIRST catalogue):
\begin{itemize}
\item Fit must converge and have an acceptable $\chi^2$ value.
\item The centre of each fitted component must be within the extent of
  the island.
\item The separation between any pair of components must be at least
  two pixels.
\item The FWHM of each component must be at least 60\% of the minimum
  FWHM of the beam.
\item The flux of each component must be positive and more than half
  the detection threshold.
\item The peak flux of each component must be less than twice the peak
  flux of the detected object and the sum of all component fluxes must
  be less than twice the total flux of the detected object.
\end{itemize}

\selavy can fit for different numbers of Gaussians, and choose the
best fit according to one of two rules. One is to look at the
reduced-$\chi^2$ value ($\chi^2/\nu$) for each fit. Here, $\nu$ is the
number of degrees of freedom in the fit, defined by the number of
pixels fitted to, $n$, and the number of parameters being fitted. In
the case of fitting all six parameters of the two-dimensional
Gaussian, a fit with $g$ Gaussians has $\nu=n-6g-1$ degrees of
freedom. Of the acceptable fits (judged according to the above
criteria), the one with the lowest $\chi^2/\nu$ is chosen as the best
fit. 

This approach has its problems, particularly for radio data that is
correlated over the size of the beam. This breaks the assumption
underlying the use of the $\chi^2$-minimisation technique, namely that
the data points be independent. While the minimisation procedure
will still work, the comparison of different fits with different
numbers of Gaussians tends to give stronger weight to fits with more
Gaussians. 

The alternative approach is to start with a single Gaussian, then only
consider fits with more Gaussians when the fit is not acceptable
according to $\chi^2$ and RMS criteria. This way the acceptable fit
with the smallest number of Gaussians is chosen. Of course, if a
larger number of Gaussians provides a better fit than a smaller
number, but the smaller one yields an acceptable fit, the better fit
will never be realised.

\subsection{Related parameterisation}

The Gaussian fitting can also be used, in certain circumstances, to
find the spectral index and spectral curvature of components. The
imaging pipeline for ASKAP will process continuum data using the
multi-scale multi-frequency synthesis algorithm (see \citet{rau09,rau10} for
a description). This produces a series of Taylor term images,
reflecting the frequency dependence of the data. 

If one expresses the frequency dependence of a continuum source's
spectrum as 
\[
I(\nu) = I_0 \left(\frac{\nu}{\nu_0}\right)^{(\alpha + \beta \log(\nu/\nu_0)},
\]
which is a quadratic function in log-space
\[
\log(I(\nu)) = \log(I_0) +
\alpha\log\left(\frac{\nu}{\nu_0}\right) +
\beta\left[\log\left(\frac{\nu}{\nu_0}\right)\right]^2, 
\] 
(where $\nu_0$ is the reference frequency, $I_0=I(\nu_0)$, $\alpha$ is the spectral
index and $\beta$ the spectral curvature), then the Taylor expansion
about $\nu_0$ becomes
\begin{eqnarray*}
I(\nu) &= &I_0 + I_0\alpha\left(\frac{\nu-\nu_0}{\nu_0}\right) +\\
&&I_0\left[\frac{1}{2}\alpha(\alpha-1) +
\beta\right]\left(\frac{\nu-\nu_0}{\nu_0}\right)^2 + ...\\
\end{eqnarray*}
The Taylor term images that come out of the pipeline then have $I_0$,
$I_0\alpha$ and $I_0 (\frac{1}{2}\alpha(\alpha-1)+\beta)$. The
source-finding and fitting of Gaussian components is performed on the
first (Taylor-0) image, and then each component has a $\alpha$ and
$\beta$ value measured. This is done by taking the component fitted to
$I_0$, keeping all parameters fixed except the peak flux, and fitting
it to the $I_0\alpha$ (Taylor-1) image. The ratio of the integrated
fluxes of this and the original component provides the spectral index
for the component. Similarly, the spectral curvature can be calculated
by fitting in the Taylor-2 image.

\section{Performance Considerations}

\begin{table*}[ht]
  \begin{center}
    \caption{Execution times on \textit{epic} for different
      configurations of two-dimensional source finding.}
    \label{tab-times-cont}
    \begin{tabular}{ll|rrl|rrl|rrl}
      \hline
$n\sigma^a$ &Config$^b$ &\multicolumn{3}{c|}{Basic Search}
&\multicolumn{3}{c|}{Reconstruction} &\multicolumn{3}{c}{Variable
  Threshold} \\
 &  &No fit$^c$ & Fit$^d$ &\# src$^e$  &No fit & Fit &\# src  &No fit &Fit  &\# src \\  
 \hline 
10	&$1\times1$	& 2.9	& 37.7	& 1484	& 87.0	&143.8	& 2358	&5732.6	&5848.0	&1367\\  	  
	&$2\times2$	& 1.8	& 11.5	& 1462	& 20.2	& 40.5	& 2327	&1358.3	&1363.9	&1367 \\
	&$3\times3$	& 1.6	&  7.2	& 1434	&  9.9	& 20.3	& 2300	& 637.8	& 641.8	&1367 \\
	&$3\times5$	& 2.2	&  6.0	& 1429	&  6.4	& 15.4	& 2301	& 382.6	& 390.4	&1367 \\
	&$5\times7$	& 1.4	&  5.3	& 1421	&  4.1	& 10.1	& 2282	& 165.6	& 178.7	&1367 \\
\hline
5	&$1\times1$	& 4.9	& 85.6	& 3547	& 90.7	&344.0	& 7119	&5468.6	&5534.3	&3133 \\
	&$2\times2$	& 3.1	& 32.0	& 3510	& 23.4	&114.1	& 6927 	&1372.4	&1379.6	&3133 \\
	&$3\times3$	& 2.7	& 21.6	& 3438	& 12.7	& 93.3	& 6703	& 635.0	& 649.9	&3133 \\
	&$3\times5$	& 2.5	& 20.8	& 3405	&  8.5	& 77.6	& 6673	& 383.2	& 403.8	&3133 \\
	&$5\times7$	& 2.8	& 15.9	& 3370	&  5.7	& 56.1	& 6576	& 165.2	& 213.3	&3133 \\
\hline
3	&$1\times1$	&14.6	&324.3	&10464	&125.2	&996.0	&20926	&5515.1	&5759.2	&8200 \\
	&$2\times2$	& 8.8	&124.9	&10309	& 39.9	&470.3     &20372	&1416.7	&1463.2	&8200 \\
	&$3\times3$ 	& 7.3	&106.9	& 9966	& 25.5	&265.6	&19738	& 641.0	& 698.1	&8200 \\
	&$3\times5$	& 7.1	& 84.4	& 9852	& 21.1	&263.1	&19767  & 390.7	& 474.6	&8200 \\
	&$5\times7$	& 7.8	& 79.0	& 9701	& 18.3	&217.4	&19392	& 169.8	& 374.5	&8200 \\
\hline
\end{tabular}
\medskip\\
$^a$ The detection threshold as a signal-to-noise ratio.\\
$^b$ The distributed worker arrangement -- the number of subdivisions
in the x- and y-directions.\\
$^c$ The time (in seconds) taken to perform the search, without doing
the 2D fitting.\\
$^d$ The time (in seconds) taken to perform the search and fit each
source.\\
$^e$ The number of sources found in the search.\\
\end{center}
\end{table*}

\begin{table*}[ht]
  \begin{center}
    \caption{Execution times on \textit{epic} for different
      configurations of three dimensional source finding.}
    \label{tab-times-sl}
    \begin{tabular}{ll|rrl|rrl|rrl}
      \hline
$n\sigma^a$ &Config$^b$ &\multicolumn{3}{c|}{Basic Search}
&\multicolumn{3}{c|}{Reconstruction} &\multicolumn{3}{c}{Variable
  Threshold} \\
 &  &Av.$^c$ & Median$^d$ &\# src$^e$  &Av. & Median &\# src  &Av. &Median  &\# src \\  
 \hline 
10	&$5\times3\times4$	&207.9	&216.2	&13    	&594.4	&593.4	&14	&1322.5	&1297.2	&11 \\
	&$7\times6\times4$	&190.9	&179.0	&13    	&53.0	&53.0	&14	&480.6	&466.5	&11 \\
	&$5\times7\times13$	&    3.6	&    3.6	&13     	&22.1	&22.0	&14	&175.7	&163.7	&11  \\
5	&$5\times3\times4$	&177.5	&179.7	&434	&737.4	&743.0	&37	&1346.8	&1377.2	&271\\
	&$7\times6\times4$	&166.0	&  95.0	&435	&94.3	&69.0	&37	&1372.0	&1261.2	&273\\
	&$5\times7\times13$	&    4.4	&    4.4	&421   	&22.3	&22.2	&34	&398.4	&396.3	&274  \\
\hline
\end{tabular}
\medskip\\
$^a$ The detection threshold as a signal-to-noise ratio.\\
$^b$ The distributed worker arrangement -- the number of subdivisions
in the x-, y- and z-directions.\\
$^c$ The average time (in seconds) of three trials taken to perform the search. \\
$^d$ The median time (in seconds) of three trials taken to perform the search.\\
$^e$ The number of sources found in the search.\\
\end{center}
\end{table*}

In this section we briefly consider the impact on processing time of
the different features discussed in the previous three sections. While
processing time can be very dependent on the system being used, we
conducted these trials on \textit{epic}, being the machine that will
be used for the BETA processing. 

We present the results of two sets of tests. One is run on a
two-dimensional image from the 2011 ASKAP simulations, being the
$4096\times4096$ central portion of the Stokes I Taylor-0 continuum
image (this just excludes the outer regions of the full
$5500\times5500$ image, where there is minimal coverage from the PAF
beams). The second set is run on the continuum-free spectral-line
simulation, using the trimmed cube of size $1248\times1248\times4096$
(or $\sim24$GB). Both sets were searched to 10 and 5 $\sigma$, with
detections grown out to $3\sigma$. The continuum image was also
searched to $3\sigma$ -- doing this on the spectral-line cube provides
far too many false detections to be feasible.

This search was done in a number of different ways. Firstly, the
processing was split over a different number of CPU cores. The
continuum image used either 1, 4, 9, 15 or 35 worker processes, while
the spectral-line cube used a smaller set of 60, 168 or 455. For each
arrangement, we consider three search types: a ``basic'' search, with
no pre-processing and a single threshold for the entire field; a
2-dimensional or 1-dimensional wavelet reconstruction (for the
continuum and spectral-line cases respectively) followed by a search
with a single threshold; and a search using the variable threshold
technique from Sec.~\ref{sec-varthresh} (using ``boxes'' in two or one
dimensions for the continuum and spectral-line cases
respectively). For each case in the two-dimensional tests, we run the
search both with and without Gaussian fitting to the detected
sources. The wavelet reconstruction is included to give a feel for
what additional time is required for pre-processing -- this is
completely separate work from the searching and/or fitting, and so the
excess time in the reconstruction case for otherwise identical
searches provides the time spent doing the reconstruction. Relative
scaling of different \duchamp pre-processing techniques can be found
in \citet{whiting12}.

The results are compiled in Table~\ref{tab-times-cont} and
Table~\ref{tab-times-sl}, where we quote the times taken by various
search types, as well as the number of sources found in each. These
searches have not been optimised in any way, and so the number of
sources do not represent the results of a complete or reliable
search. Note also that these tests were run on a shared system, and so
some variability in the timing is expected. The values are averages
over several runs, but even so some variation from the expected trends
can be seen when the durations are small. For the three-dimensional
results, we see considerable variability between the different runs
(communication delays may play a part here), and so quote the middle
value of the three times as well.

This variability aside, the first thing to note is that the search
times are fairly small, certainly compared to the time required to do
the imaging (which will be several hours at least). The exception here
are the single-node (``$1\times1$'') 2D searches with the variable
threshold. This is because the median calculations, in a box of size
$101\times101$ in this case, for each pixel in the image is quite
computationally intensive. However, we clearly see that with even a
modest amount of distributed processing allows the image to be done in
under 10 minutes. And note that the results are the same for all
distributed processing arrangements, since whether each pixel is
detected or not depends only on the box surrounding them. This does
not apply to the other searches, as the threshold is determined from
the statistics, which are calculated in the distributed fashion.

The additional time for the fitting is governed primarily by the
number of sources to be fitted. Note that these are from a $\sim10$
square degree image resulting from a simulated 12-hour integration
with the full ASKAP array. For the distributed cases, we see an
average of a few to 10 milliseconds per source, which in absolute
terms does not prove to be a big additional cost, and allows the
consideration of additional, more complex, fitting algorithms.

The spectral-line tests demonstrate the need to perform
three-dimensional source-extraction on a distributed system. We have
been able to search a 24GB cube, often in a matter of minutes or less,
and we see that increasing the number of available processors does
lead to faster execution. 

These results are encouraging for considering processing pipelines, as
they allow a lot of flexibility in algorithmic approaches
(particularly for two-dimensional searches), without impinging on the
time available to run the entire pipeline. This provides leeway in
designing source-finding algorithms that can operate within the
pipeline environment yet still deliver results appropriate for
different science cases. We discuss in the next section the processes
governing the incorporation of new algorithms into the pipeline.

\section{The Development Process and Community Involvement}
\label{sec-devel}

The ASKAP telescope is expected to have at least 75\% of its first
five years of operation devoted to Survey Science Projects (SSPs),
each requiring at least 1,500 hours of observing time. Ten such
projects have been selected to participate in a Design Study, where
the detailed scientific and technical aspects of their survey,
including the processing that is required, will be developed.

To assist communication both between different SSPs and between SSPs
and the ASKAP team, working groups were established in a small number
of key areas, one of which being source finding. A large part of the
technical work of the Design Study has been the investigation of
source-finding techniques, with the aim of providing recommendations
to the ASKAP computing team on the capabilities of the source-finding
pipeline. 

Since the prototype ASKAP pipeline is built on the \duchamp library,
the \duchamp package has formed the basis of much of the testing, as
can be seen in numerous papers in this issue
\citep{allison12,popping12,westmeier12,westerlund12,walsh12}.

However, such testing does not capture the new features implemented in
\selavy. To facilitate testing of these features, \selavy access was
provided as a service rather than an installable software
package. This service enables the Survey Science Team members to
access both the software and a modest size compute cluster provided by
CSIRO. This service is delivered via a script interface enabling
uploading of images, submission of source-finding jobs, and the
retrieval of results. This provides a mechanism to test the new
features described herein and evaluate whether they are appropriate
for the relevant science case. At least two papers
\citep{huynh12,hancock12} have made use of this service to test the
continuum image processing of \selavy.

This testing process is partly designed to allow the science teams to
develop algorithms that are either missing from the current design of
the ASKAP source-finder, or do not work to the level required by the
science. We have instituted a process whereby, once such algorithms
have been identified, they can be provided to ASKAP computing for
possible inclusion in the pipeline prior to ASKAP or BETA
observations. In this way we aim to provide a source-finder that will
have all features required by the various science cases.

New features that are currently planned to be implemented (many of
which are detailed in papers in this issue) include:
\begin{itemize}
\item Optimal extraction of spectra around detected continuum sources,
  for the purposes of further processing and analysis (such as
  one-dimensional searches, or rotation measure synthesis).
\item Mask optimisation routines, to find the optimal mask for an
  extended object, particularly in three dimensions. This will address
  the issues with the measurement of integrated flux identified by
  \citet{westmeier12}.
\item An alternate wavelet reconstruction algorithm, the 2D-1D
  algorithm \citep{floer12}, that allows the treatment of the spectral
  axis differently to the spatial axes.
\item An alternative searching technique for one-dimensional spectra
  that applies Bayesian Monte Carlo methods to detect absorption lines
  \citep{allison12}.
\item Alternative Gaussian fitting algorithms, such as those used in
  the Aegean source finder \citep{hancock12}.
\end{itemize}

The current plan is to have as many of these features as possible
available within the ASKAP pipeline in time for science observations
with BETA. Their performance will be evaluated during the
commissioning phase to plan what further work is required for
ASKAP-scale processing. We expect that algorithm development will
continue within the Survey Science Teams, and anticipate further input
as ASKAP operations approach. 

\section{Summary and Future Work}

We have presented the key algorithmic developments that have gone into
\selavy, the prototype ASKAP source-finder. These features, including
distributed processing, variable threshold determination and
two-dimensional Gaussian profile fitting, have been implemented in a
prototype system that has been made available to the ASKAP Survey
Science Teams for testing purposes.

The development of the \selavy source-finder is continuing as we move
closer to ASKAP operations. Several Science Teams have provided
feedback and specifications for additional or refined algorithms,
covering pre-processing, source extraction and parameterisation and
addressing some of the issues identified in this paper. At time of
writing, we are incorporating these algorithms into \selavy, with the
aim of providing a more fully-fledged source-finder in time for BETA
observations.

\section*{Acknowledgments} 

We acknowledge the feedback provided by the ASKAP science teams
resulting from their testing of \duchamp and \selavy, and their
contribution to the source-finding algorithms. 

This work was supported by the iVEC@Murdoch supercomputer, Epic, and
the NCI National Facility at the ANU.

This work supports the Australian Square Kilometre Array Pathfinder,
located at the Murchison Radio-astronomy Observatory (MRO), which is
jointly funded by the Commonwealth Government of Australia and State
Government of Western Australia.  The MRO is managed by the CSIRO, who
also provide operational support to ASKAP.  We acknowledge the Wajarri
Yamatji people as the traditional owners of the Observatory site.

\end{document}